\documentclass[a4paper,11pt]{article}
\usepackage{jinstpub} % for details on the use of the package, please see the JINST-author-manual
 \usepackage{subfigure}

\title{\boldmath Research on trigger technology of MRPC TOF-PET system and imaging results of $^{22}$Na radioactive source}

 \author[a]{Jianing Liu,}
 \author[b,c]{Yuelei Ma,}
 \author[a]{Ziyang Chen,}
 \author[b,c]{Zhenyan Li,}
 \author[a,1]{Yi Wang,\note{Corresponding author.}}
 \author[a]{Baohong Guo,}
 \author[a]{Dong Han,}
 \author[a]{and Yuanjing Li}
 \affiliation[a]{Key Laboratory of Particle and Radiation Imaging, Department of Engineering Physics, Tsinghua University, Beijing 100084, China}
 \affiliation[b]{State Key Laboratory of Particle Detection and Electronics, University of Science and Technology of China, Hefei 230026, China}
 \affiliation[c]{Modern Physics Department, University of Science and Technology of China, Hefei 230026, China}

\emailAdd{liu-jn20@mails.tsinghua.edu.cn}

\abstract{This study focuses on developing a self-triggered data acquisition system and a noise reduction algorithm for the Multi-gap Resistive Plate Chamber (MRPC) Time-of-Flight Positron Emission Tomography (TOF-PET) system. The system integrates a fast front-end amplifier, a waveform digitization module based on the DRS4 chip, and an efficient noise reduction algorithm to address challenges such as high noise trigger rates and precise gamma-ray detection. The proposed self-triggered system, through threshold discrimination, coincidence logic, and continuous oscillation check, reduces the noise trigger rate to 0.004 Hz. Experimental results show that the system accurately localizes and images the $^{22}$Na radioactive source, and has a good time resolution of 162 ps FWHM for 0.511 MeV gamma rays.}


\begin{document}
\maketitle
\flushbottom

\section{Introduction}
\label{sec:intro}

The concept of Multi-gap Resistive Plate Chamber (MRPC) \cite{zeballos1996new} Time of Flight (TOF) \cite{fonte2000new} Positron Emission Tomography (PET) is based on the detection principle of convert plates \cite{bateman1984rutherford}. It takes advantage of the natural layered structure of MRPCs, their simple and low cost design, excellent time resolution \cite{liu2024very}, and remarkable position accuracy \cite{blanco2012toftracker}. These features make MRPC TOF-PET particularly appealing for detailed imaging of small animals and high-sensitivity, whole-body TOF-PET imaging for humans. The Fonte team demonstrated imaging results using needle and planar $^{22}$Na sources, with the acquired data reconstructed through the Maximum Likelihood Expectation Maximization (MLEM) algorithm, achieving a stable and impressive resolution of 0.4 mm FWHM \cite{martins2014towards}. The Fonte team also developed a brain imaging-specialized Resistive Plate Chamber (RPC)-PET prototype system, achieving a reconstructed image resolution at the sub-millimetric level \cite{fonte2023rpc}. Blanco systematically analyzed the impact of the key structural parameters of RPC-PET on efficiency and imaging accuracy by simulations and experiments \cite{blanco2009efficiency}. 

MRPCs are often used as TOF systems with data acquisition relying on scintillators for external trigger \cite{wang2016performance} in high-energy physics experiments. However, MRPC TOF-PET systems cannot use scintillators to trigger gamma ray detection. A self-triggered data acquisition system is needed to detect and record events of interest. High-performance commercial TOF-PET devices utilize tens of thousands of electronics channels to achieve high sensitivity and resolution \cite{spencer2021performance}. The high noise trigger rate poses a significant challenge. The noise sources originate from electronic noise, external electromagnetic induction noise, and cosmic ray radiation background. This paper presented a self-triggered data acquisition system based on fast front-end amplifiers and waveform acquisition, along with an efficient noise reduction algorithm, aiming to enhance the accuracy of MRPC TOF-PET in detecting gamma rays. Experimental results show that the Noise Trigger Rate (NTR) of data acquisition using this self-triggered data acquisition system is significantly reduced, and the positioning and imaging results of the $^{22}$Na radiation source are accurate.

\section{System setup and data acquisition}

\begin{figure}[h]
\centering
\includegraphics[width=0.96\textwidth]{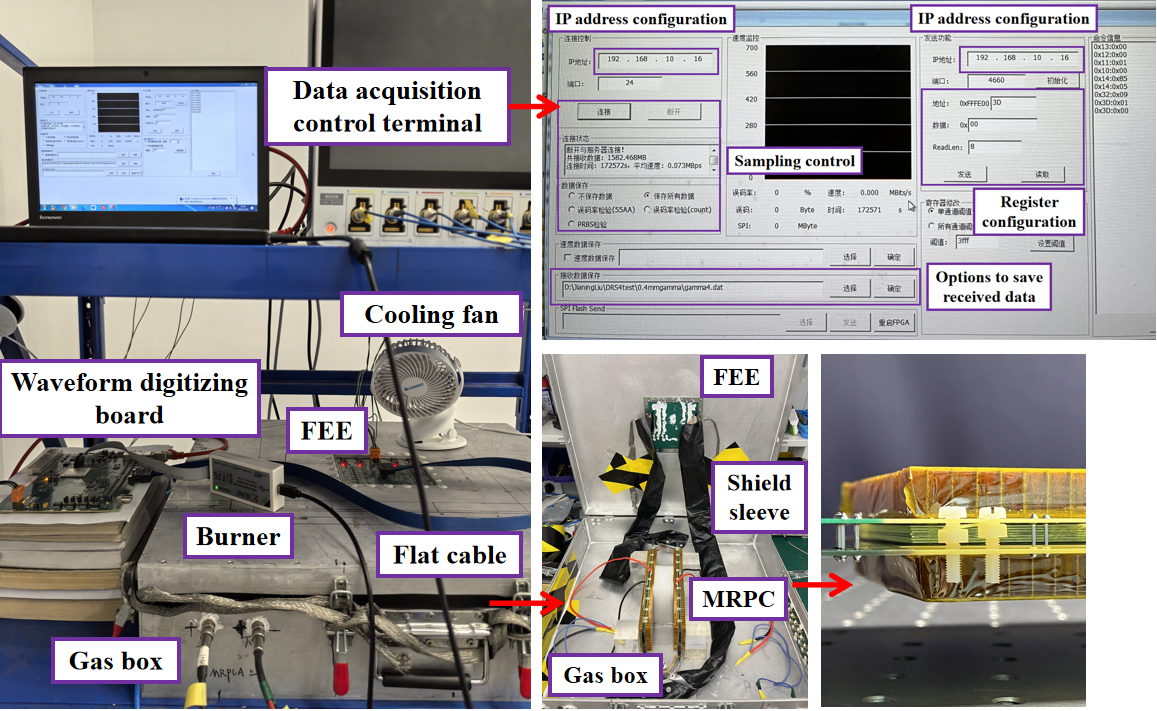}
\caption{Test system diagram.\label{fig:1}}
\end{figure}

\begin{figure}[htbp]
\centering
\includegraphics[width=0.8\textwidth]{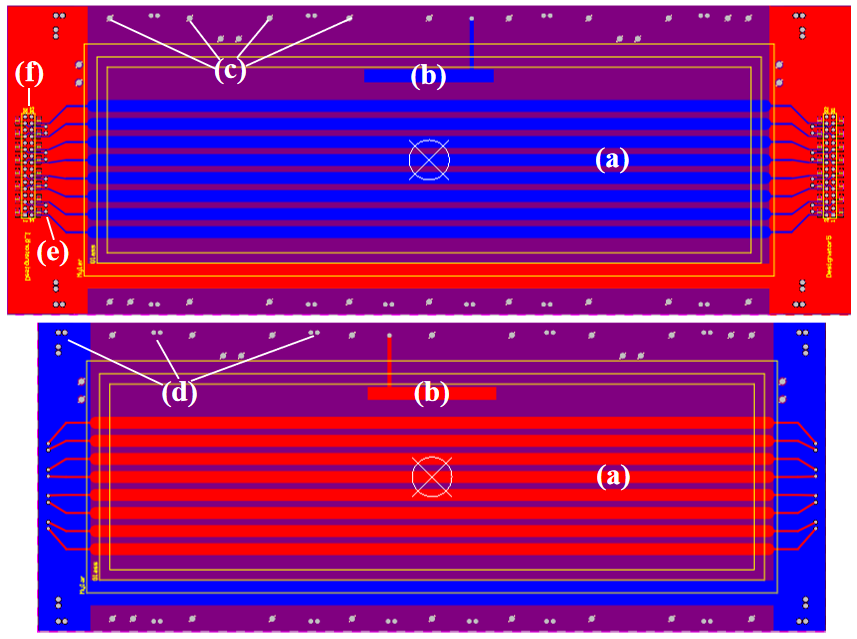}
\caption{The design of the Printed Circuit Board (PCB) for the MRPC prototype. The labeled components in the figure are as follows: (a) Eight sets of readout strips, each with a width of 5 mm and a spacing of 2 mm; (b) High-voltage copper foil; (c) Holes for installing nylon posts used for winding fishing lines; (d) Mechanical fixing holes for soldering dual-row pins; (e) Signal transmission pinholes between the top and bottom PCB; (f) Differential signal output vias.\label{fig:2}}
\end{figure}

The self-triggered system primarily consists of MRPC detectors, fast front-end amplifiers, a waveform digitization module, and a data acquisition control terminal. The system setup is shown in figure~\ref{fig:1}. The MRPC prototype has 8 gas gaps with 0.128 mm gas gap thickness and 0.4 mm resistive plate thickness. The resistive material in this detector is glass, not a high-Z material. The active area of the detector is 264×56 mm\textsuperscript{2}. Figure~\ref{fig:2} shows the design of the Printed Circuit Board (PCB) for the MRPC prototype. Signals from the readout strips on the bottom PCB are transmitted to the top PCB through signal vias (e) and soldered pins. This design ensures that the transmission paths for the positive and negative polarity signals of the MRPC are identical. The differential signals generated by the MRPCs are then routed through differential signal output vias (f) and twisted-pair cables to the front-end electronics (FEE). 

\begin{figure}[htbp]
\centering
\subfigure[]{
\label{fig：subfig_n}
\includegraphics[width=.35\textwidth]{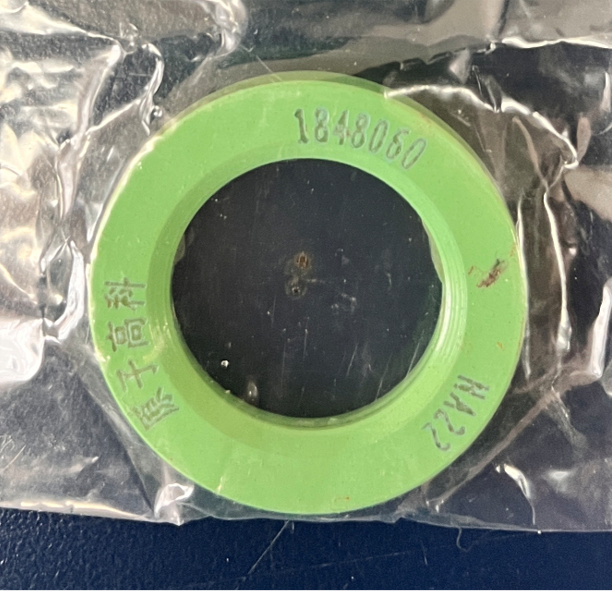}
}
\subfigure[]{
\label{fig：subfig_o}
\includegraphics[width=.55\textwidth]{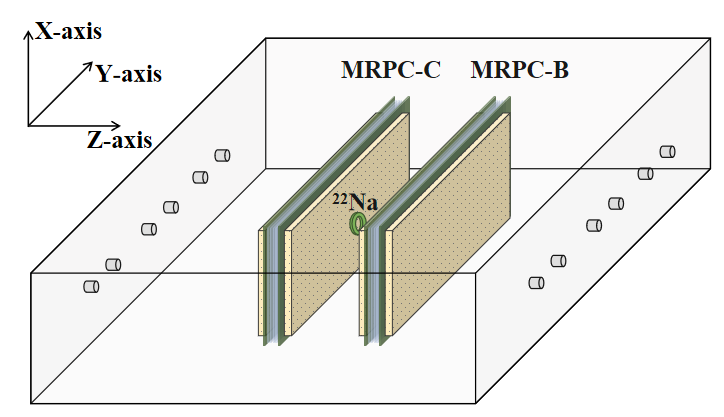}
}
\caption{(a) Physical diagram of the $^{22}$Na radioactive source used in the Gamma test. (b) Schematic diagram of the definition of the X, Y, and Z-axis coordinates.\label{fig:8}}
\end{figure}

The physical diagram of the $^{22}$Na radioactive source used in the Gamma test is shown in figure \ref{fig:8}(a). The diameter of the active area is 3 mm, and the activity is \(1.57 \times 10^{5}\) Bq. The definitions of the X, Y, and Z axes are shown in figure \ref{fig:8}(b). The X-axis is defined as the direction perpendicular to the readout strips. The zero point of the X-axis is set at the midpoint of the 6th readout strip from the top down. The Y-axis is defined as the direction along the readout strips. The length of the readout strip is 264 mm, and the zero point of the Y-axis is set at the midpoint along the direction of the readout strip. The Z-axis is defined as the direction perpendicular to the MRPC detector plane. The zero point of the Z-axis is set at the center of the MRPC-C detector. The $^{22}$Na source is placed between MRPC-C and MRPC-B, 32 mm away from the MRPC-C detector and 40 mm away from the MRPC-B detector. The position coordinates of the $^{22}$Na placement are (16 mm, 4.5 mm, 32 mm).

The following hardware-level noise reduction optimizations have been implemented for signal readout: Differential readout is adopted, and input interfaces are equipped with ground vias. These ground vias help optimize the return path of signals and reduce the influence of external noise. Both sides of the PCB are copper-clad and connected to the ground. The front-end electronics are embedded in an aluminum gas box, with the MRPC detector's ground connected to the front-end electronics' ground. Signal transmission uses twisted-pair cables wrapped in aluminum foil shielding sleeves, as shown in figure~\ref{fig:1}. Utilizing twisted-pair cables to transmit signals to the electronics improves their noise immunity.
\begin{figure}[htbp]
\centering
\subfigure[]{
\label{fig：subfig_a}
\includegraphics[width=.55\textwidth]{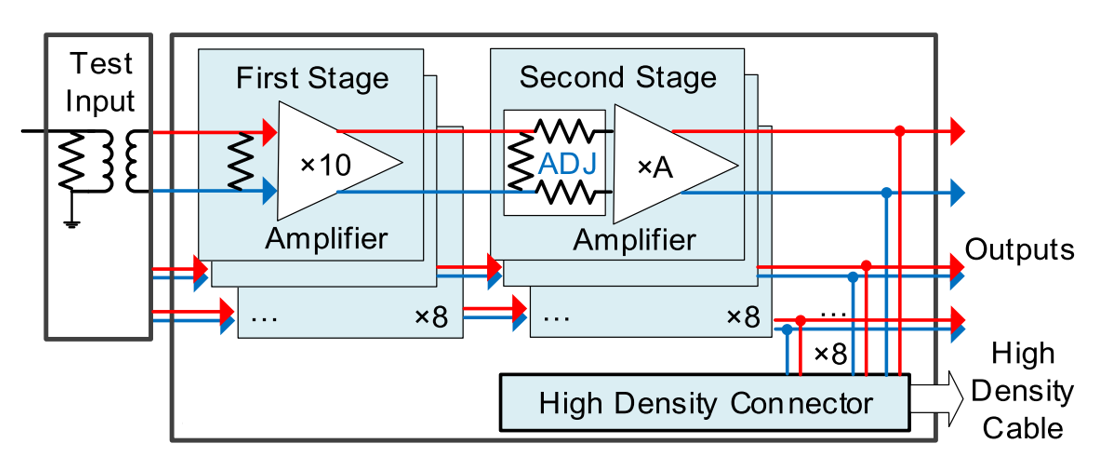}
}
\subfigure[]{
\label{fig：subfig_b}
\includegraphics[width=.32\textwidth]{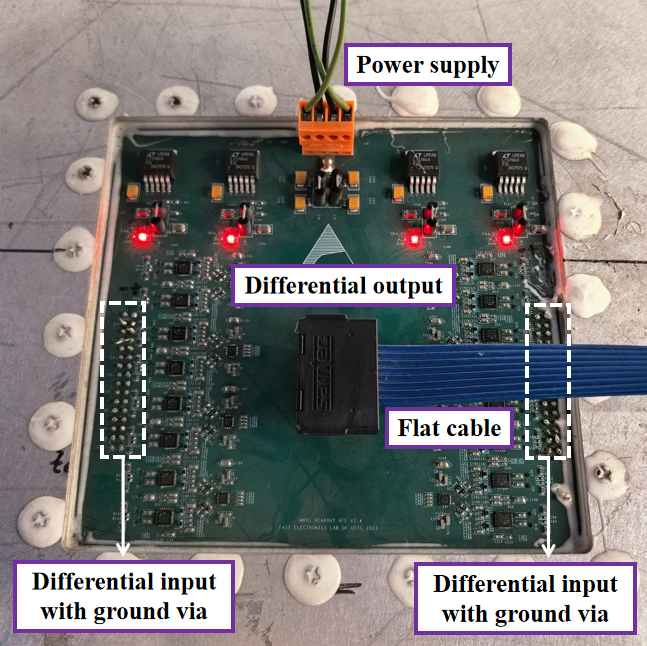}
}
\caption{Front-end amplifier module. (a) Schematic diagram, (b) physical prototype.\label{fig:3}}
\end{figure}

Our collaborator, the University of Science and Technology of China (USTC), completed the development of the fast front-end amplifier module and the waveform digitization module. Figure \ref{fig:3}(a) shows the schematic diagram of the front-end amplifier module, which adopts a two-stage cascaded amplification structure. This design ensures low noise and high bandwidth while addressing the limitation of insufficient amplification gain in single-stage solutions. The first stage consists of coupling resistors and the LTC6430-20, a 20 dB ultralow-noise, high-bandwidth radio-frequency amplifier. The second stage uses the ADL5569, a 20 dB high-bandwidth operational amplifier. A 16-channel fast front-end amplifier board, as shown in figure \ref{fig:3}(b), was fabricated. The input and output of the amplifier board are differential signals, with a configurable gain ranging from 20 to 100 times. The -3 dB bandwidth exceeds 1.4 GHz, and for input signals in the range of 5 mV to 50 mV, the amplifier time jitter is better than 4 ps RMS \cite{liu2019design}.

\begin{figure}[htbp]
\centering
\subfigure[]{
\label{fig：subfig_c}
\includegraphics[width=.55\textwidth]{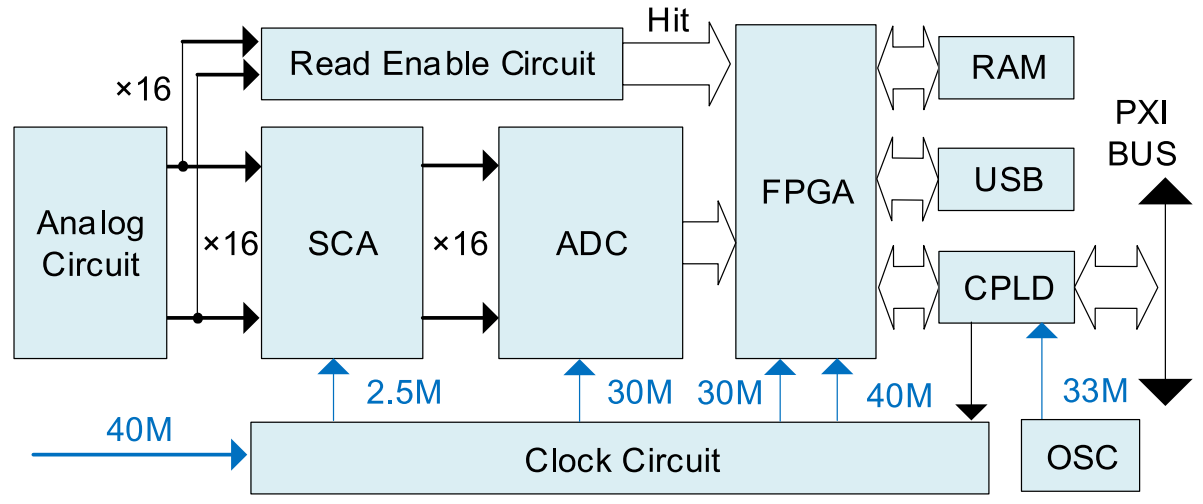}
}
\subfigure[]{
\label{fig：subfig_d}
\includegraphics[width=.32\textwidth]{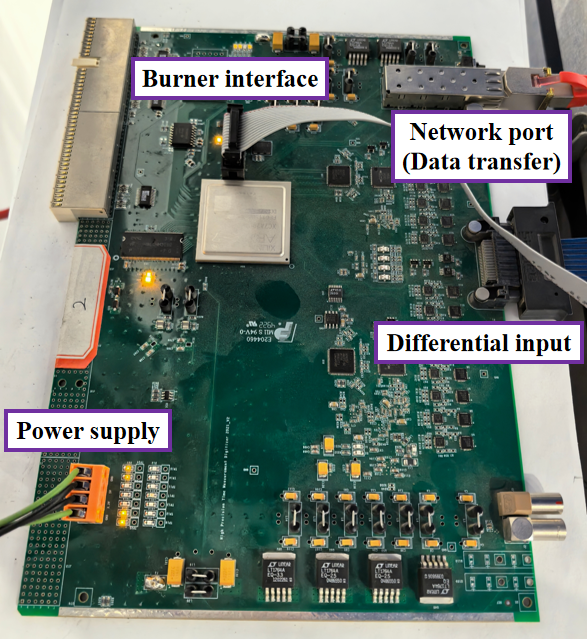}
}
\caption{Waveform digitization module. (a) Schematic diagram, (b) physical prototype.\label{fig:4}}
\end{figure}

The structural schematic and physical prototype of the waveform digitization module are shown in figure \ref{fig:4}. This module is designed based on the Domino Ring Sampler (DRS4) chip, Analog-to-Digital Converter (ADC) chip, and Field Programmable Gate Array (FPGA), and it includes analog front-end circuits, DRS4 waveform sampling circuits, analog-to-digital conversion circuits, and digital signal processing circuits. Signals from the front-end amplifier are buffered by an analog circuit to enhance the driving capacity of the sampling circuit and the readout enabling circuit. The signals are then continuously sampled and stored in the 1024 sampling capacitors of Switching Capacitor Array (SCA) chip called DRS4. 

Once the signal exceeds the threshold of the readout enabling circuit, the stored charges are sequentially read out and fed into a 14-bit ADC, which digitizes them at a speed of 30 MHz for optimal linearity. After digitization, the ADC output data are input to the FPGA for further computation. The computed data can be transferred to a PC via the PXI bus through a Complex Programmable Logic Device (CPLD). The CPLD also configures the clock circuit and FPGA logic during power-up. This high-speed, high-precision waveform digitization board features 16 channels. Testing with sine waves and high-speed pulses demonstrates that the module operates at a sampling rate of 5.12 Gsps, with a -3 dB analog bandwidth of approximately 620 MHz. Additionally, its time resolution was verified by measuring the time difference of a high-speed pulse signal split into two parts. For input signal amplitudes ranging from 100 mV to 1 V, the single-channel time precision is better than 8 ps \cite{liu2019design}.
\section{Self-triggered noise reduction algorithm}
\begin{figure}[h]
\centering
\subfigure[]{
\label{fig：subfig_e}
\includegraphics[width=.35\textwidth]{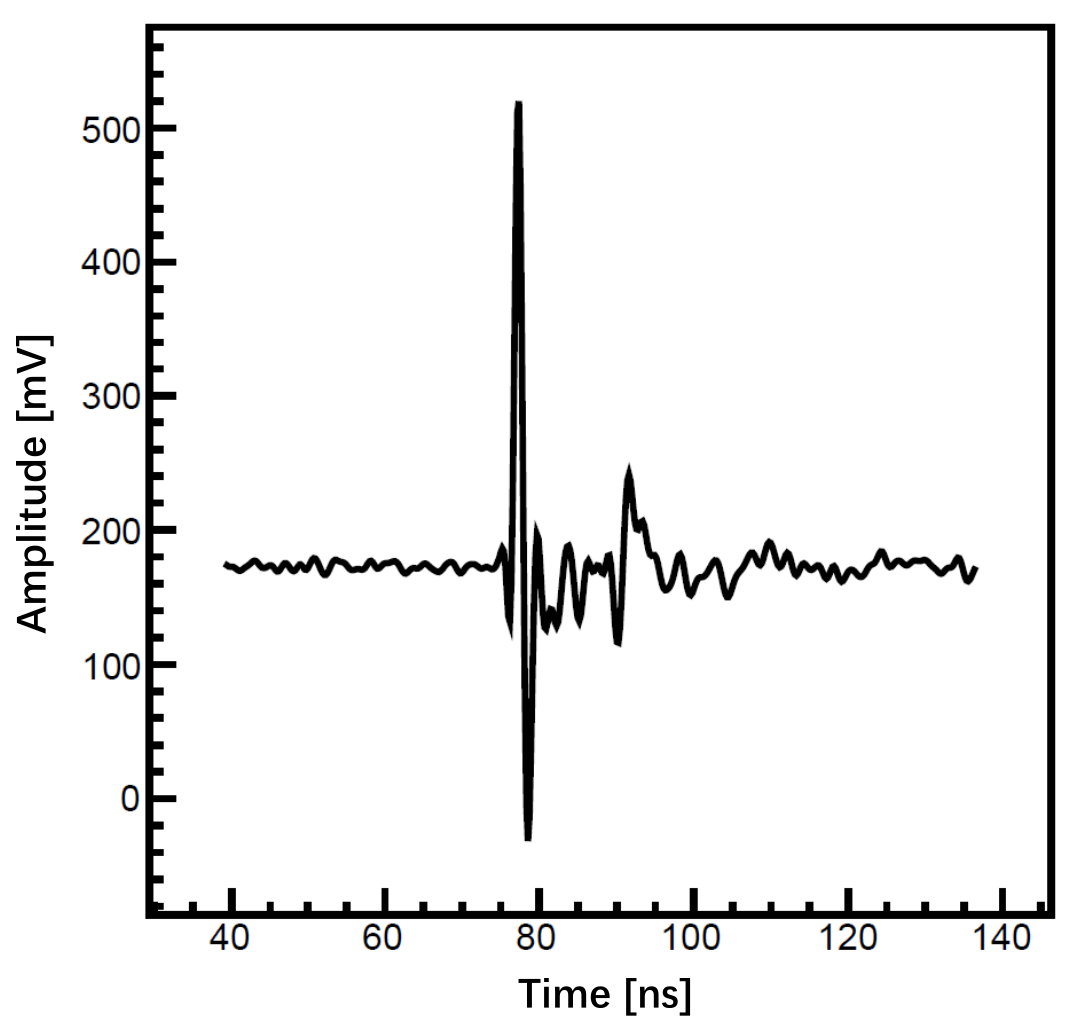}
}
\subfigure[]{
\label{fig：subfig_f}
\includegraphics[width=.35\textwidth]{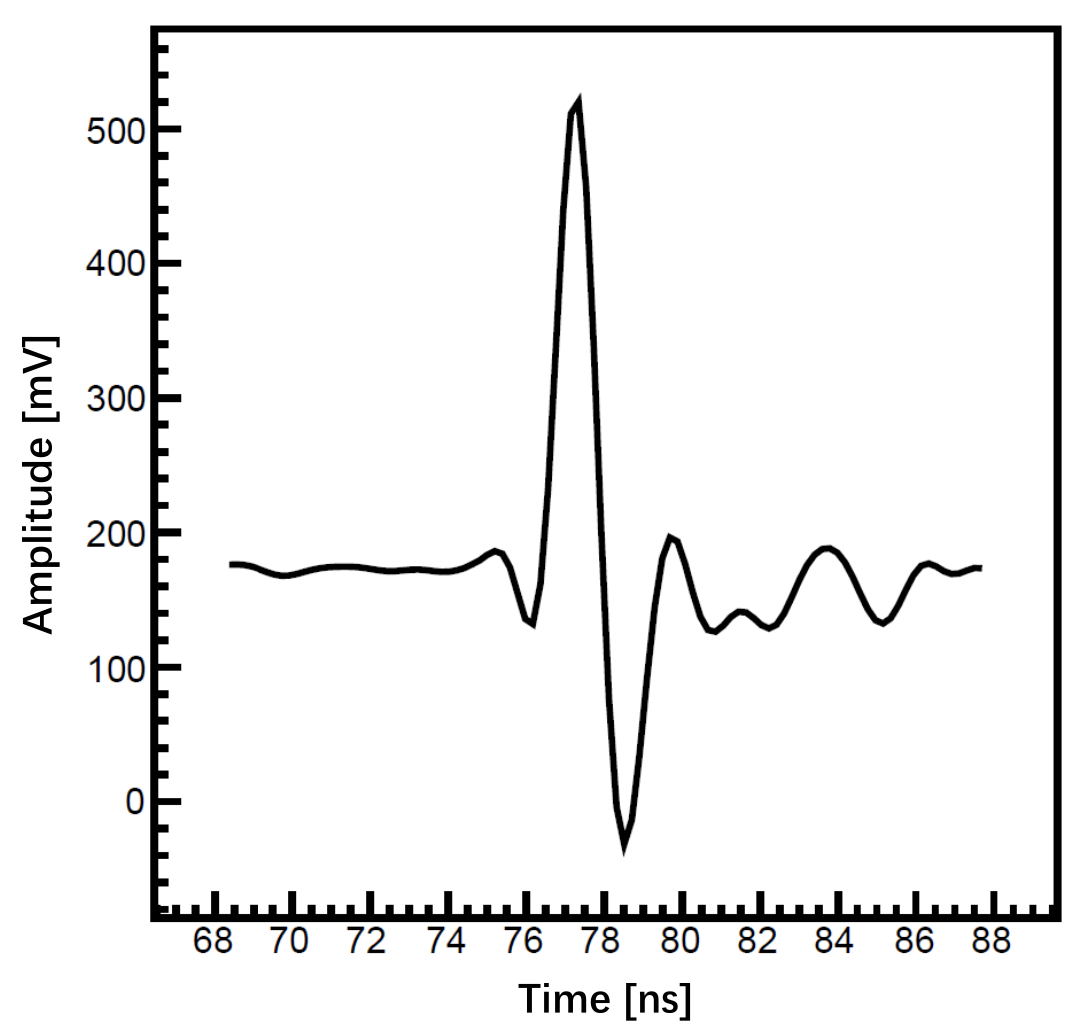}
}
\caption{The waveform generated by the MRPC detector using the fast front-end amplifier and the waveform digitization module. (a) A complete signal waveform in the experiment, (b) local amplification of the waveform.\label{fig:5}}
\end{figure}
\begin{figure}[h]
\centering
\subfigure[]{
\label{fig：subfig_g}
    \includegraphics[width=.45\textwidth]{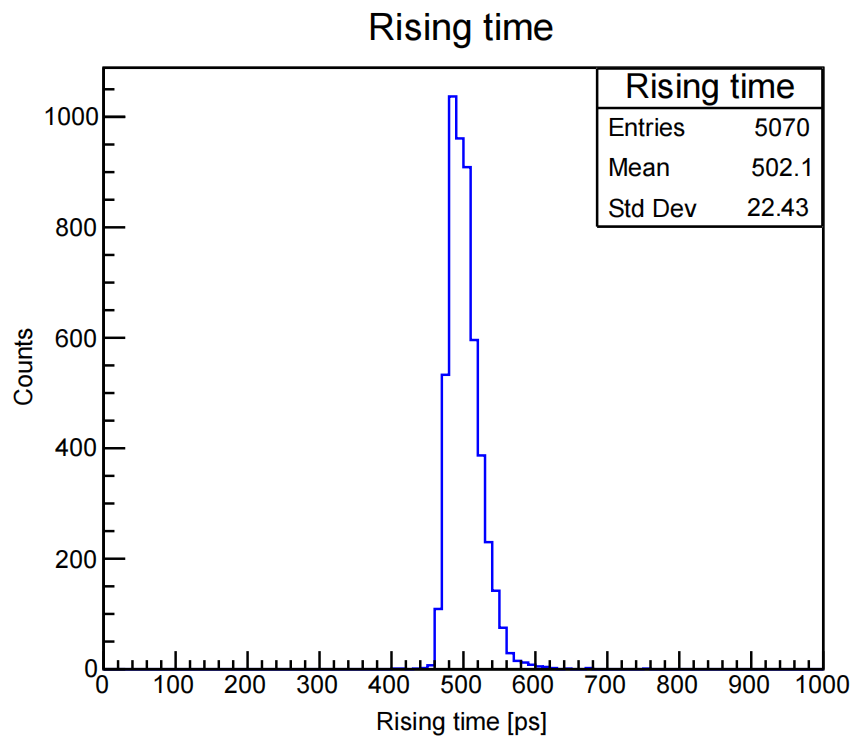}
}
\subfigure[]{
\label{fig：subfig_h}
    \includegraphics[width=.45\textwidth]{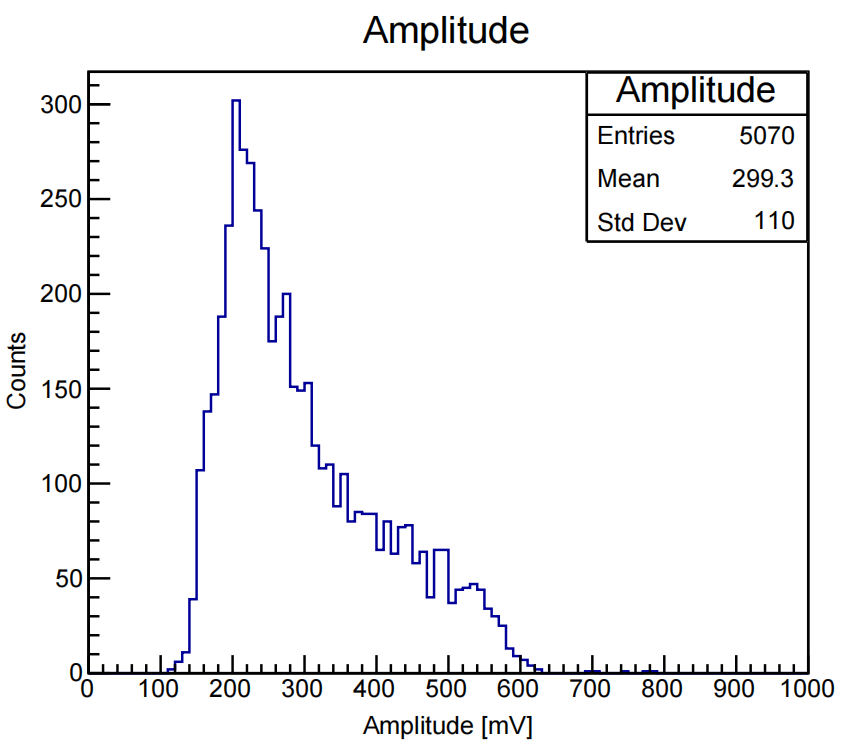}
}
\subfigure[]{
\label{fig：subfig_i}
    \includegraphics[width=.45\textwidth]{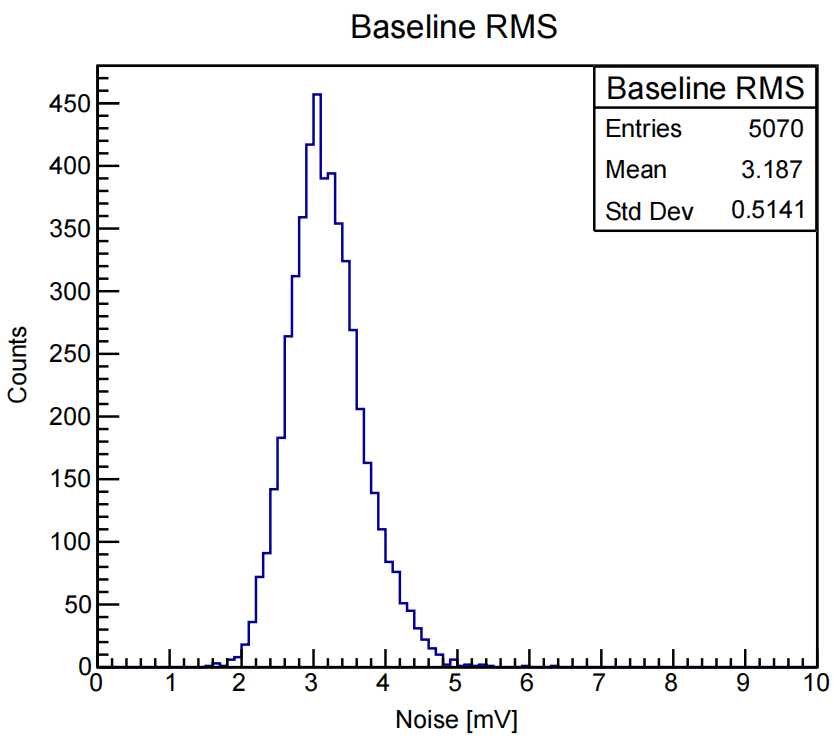}
}
\caption{The signal characteristics of the 8-gap MRPC detecting 0.511 MeV gamma rays emitted by a ${}^{22}{\rm{Na}}$ radioactive source. (a) The distribution of signal rise time, (b) the distribution of signal amplitude, (c) the distribution of baseline RMS.\label{fig:6}}
\end{figure}
\begin{figure}[h]
\centering
\subfigure[]{
\label{fig：subfig_j}
    \includegraphics[width=.45\textwidth]{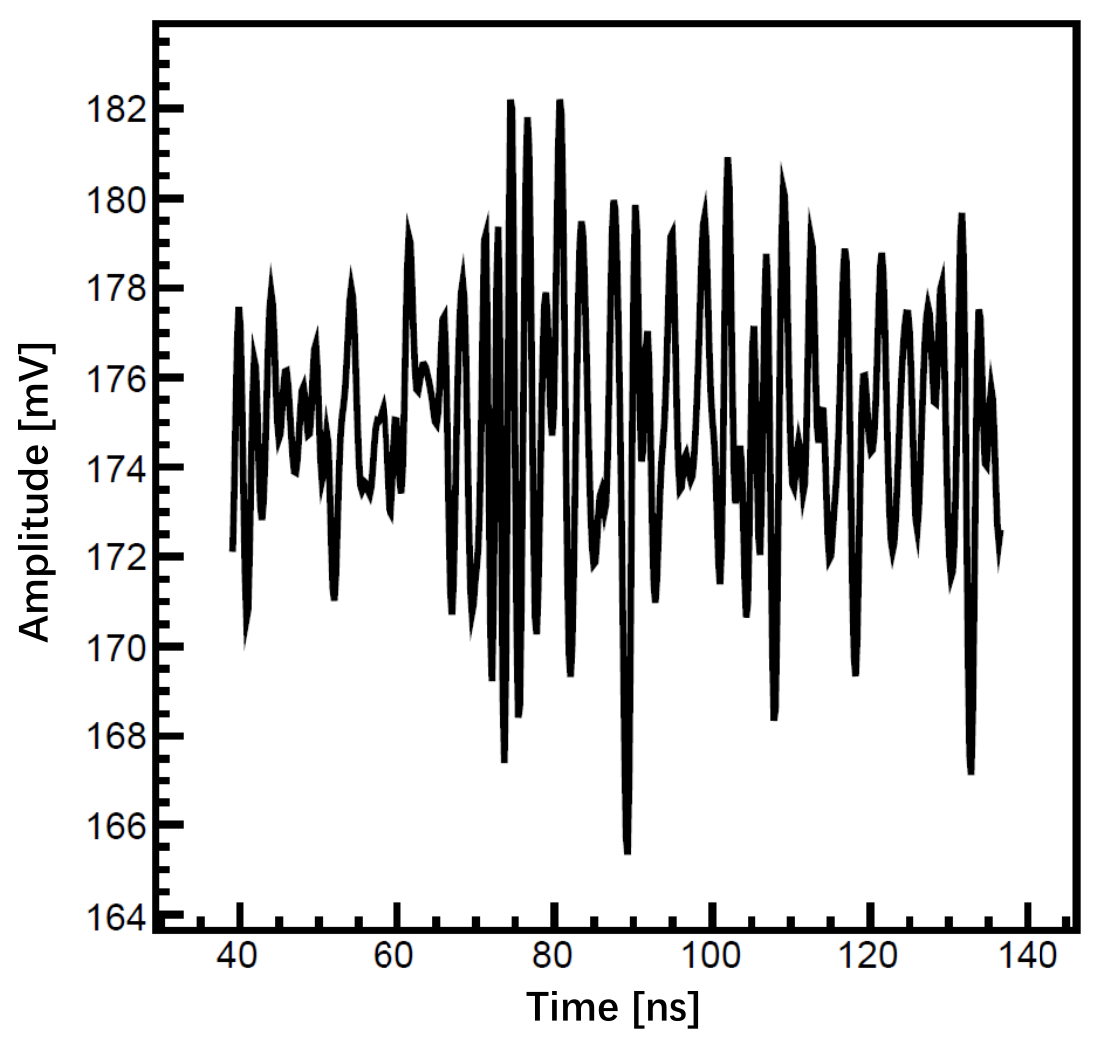}
}
\subfigure[]{
\label{fig：subfig_k}
\includegraphics[width=.45\textwidth]{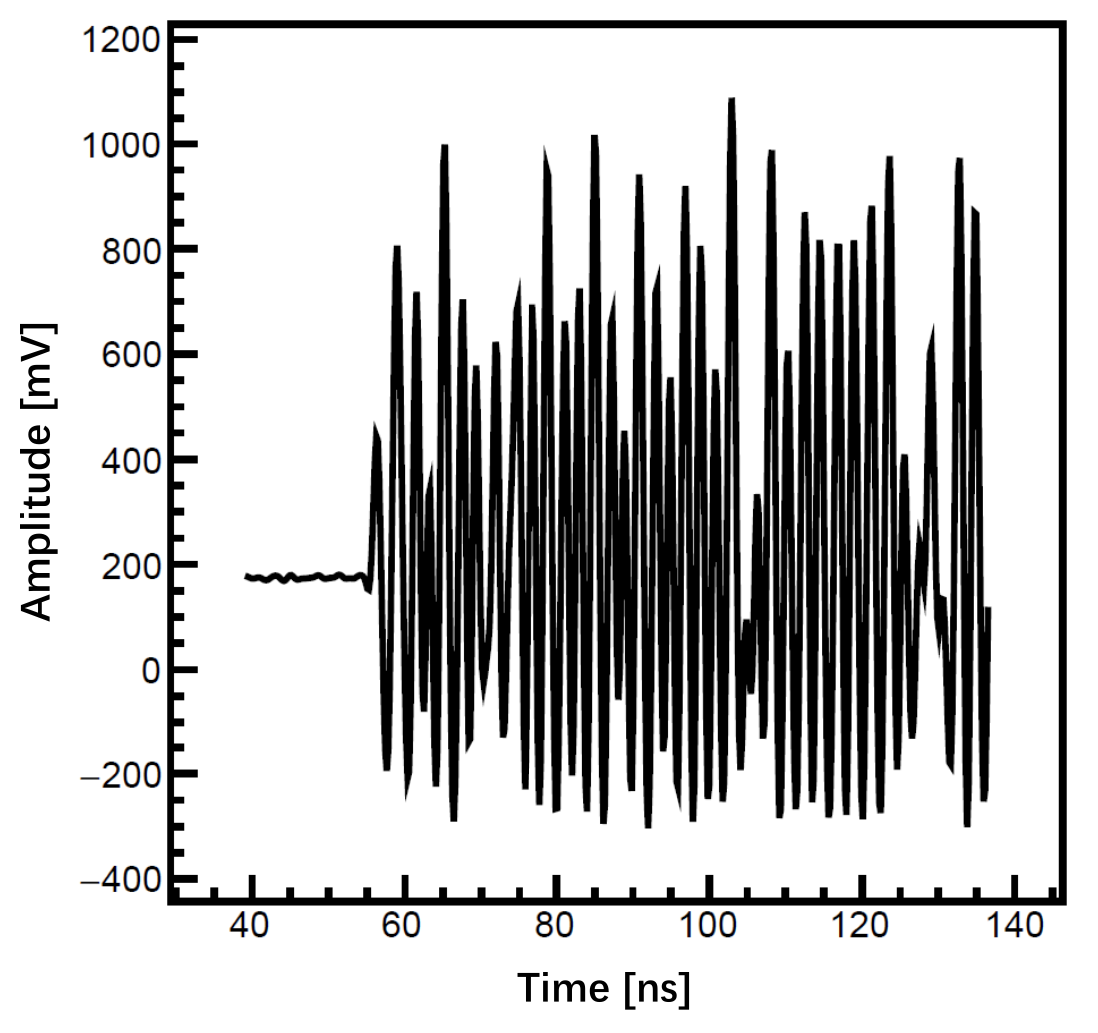}
}
\subfigure[]{
\label{fig：subfig_l}
    \includegraphics[width=.45\textwidth]{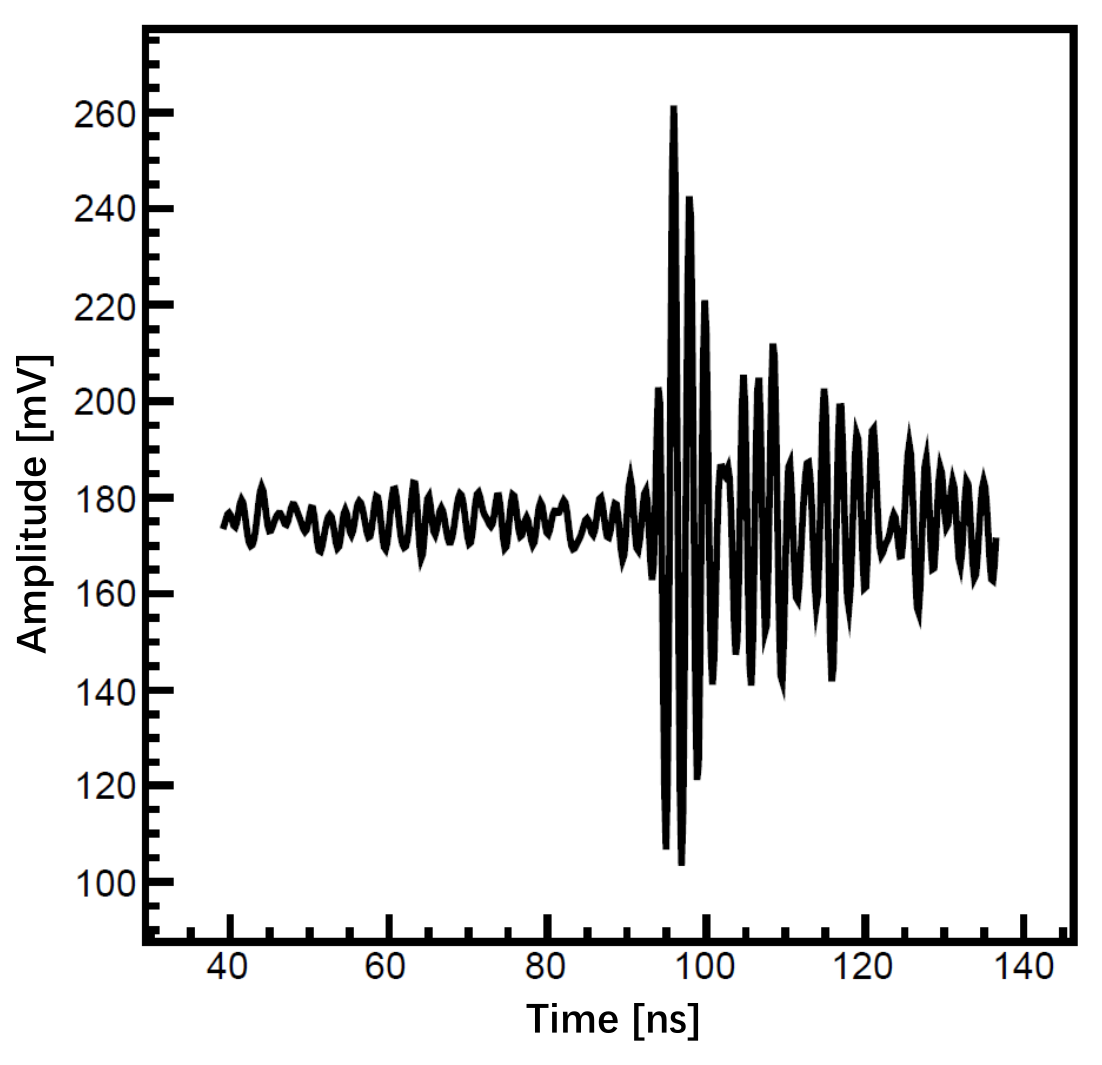}
}
\subfigure[]{
\label{fig：subfig_m}
    \includegraphics[width=.45\textwidth]{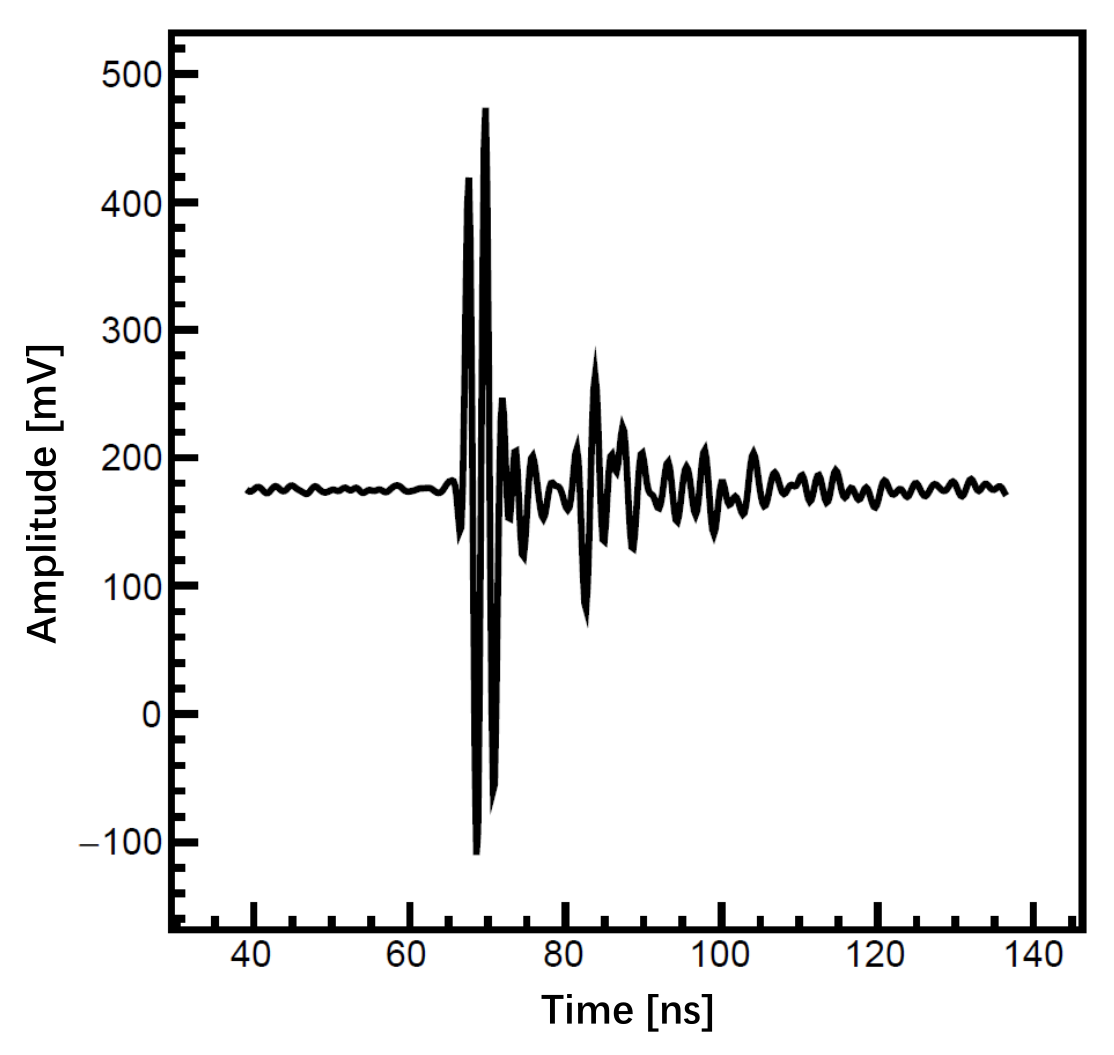}
}
\caption{Examples of several complex waveforms from the 8-gap MRPC using the waveform digitization module. The complex waveforms are noise events. The origin of these pulses are electronic noise, electromagnetic induction noise, and cosmic ray radiation background.  \label{fig:7}}
\end{figure}
The signal waveforms generated by the MRPC detector using the fast front-end amplifier and the waveform digitization module when detecting 0.511 MeV gamma rays emitted by a ${}^{22}{\rm{Na}}$ radioactive source, are shown in figure \ref{fig:5}. The rise time of the signal is defined as the time it takes for the signal to rise from 10\% to 90\% of its maximum amplitude. The rise time distribution and amplitude distribution of signals detecting 0.511 MeV gamma rays are shown in figure \ref{fig:6}. The average rise time of signals is 502.1 ps with a typical pulse width of 1–2 ns, and the average signal amplitude is 299.3 mV. 

The MRPC detector requires self-trigger to detect gamma rays. The noise sources include electronic noise, electromagnetic induction noise, and cosmic ray radiation background. Figure \ref{fig:7} shows several complex waveforms captured using the waveform digitization module. The complex waveforms are noise events. The origin of these pulses are electronic noise, electromagnetic induction noise, and cosmic ray radiation background. In figure \ref{fig:7}(a), the noise amplitude is relatively low, allowing low-amplitude noise signals to be preliminarily filtered out using threshold discrimination. Further, by applying logical coincidence checks, the system verifies whether signals in multiple channels simultaneously meet specific conditions, helping to distinguish noise signals from gamma events and reducing the system's NTR. The NTR refers to the frequency of false triggers caused by noise in the absence of valid signal inputs. It is a key metric for evaluating the noise immunity of the system. The NTR, typically expressed in triggers per second (Hz), is defined as the number of noise-induced triggers per unit time:
\begin{equation}
\label{eq:1}
\begin{aligned}
NTR = \frac{{{N_{noise}}}}{\Delta t}
\end{aligned}
\end{equation}
where $N_{noise}$ is the number of noise triggers within a certain time interval $t$. When the trigger threshold is set to 100 mV and the coincidence time window to 5 ns, the system's NTR is 0.9 Hz. 
\begin{figure}[h]
\centering
\includegraphics[width=0.45\textwidth]{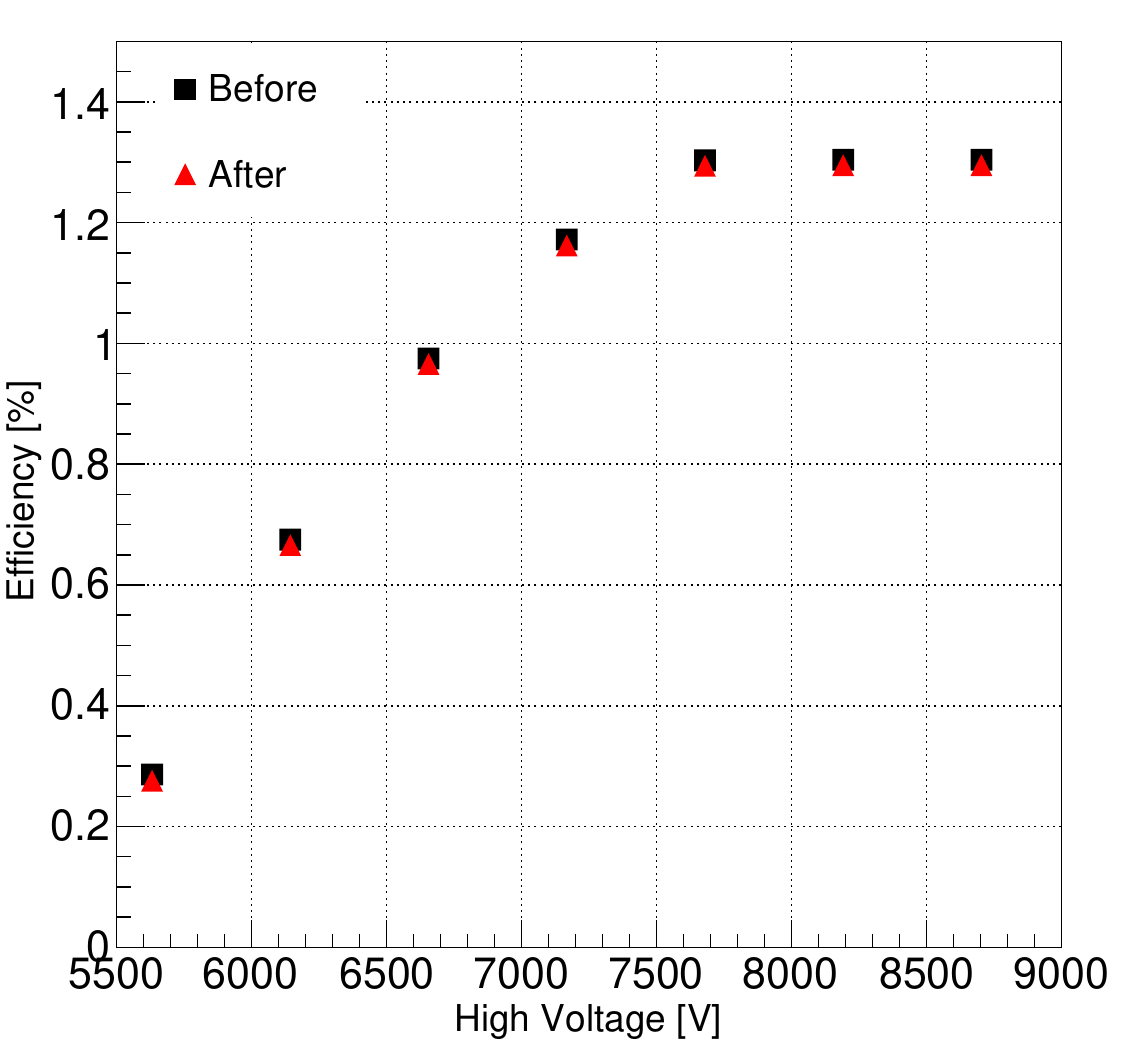}
\caption{The detection efficiency of the MRPC prototype as a function of High Voltage, before and after the continuous oscillation check. \label{fig:13}}
\end{figure}

To further reduce the system's NTR and minimize the waste of data storage and processing resources, we introduced a continuous oscillation check during the data acquisition process. The continuous oscillation check is based on the characteristic that noise signals often exhibit high-frequency, continuous oscillations, as shown in figure \ref{fig:7}(b), whereas gamma event signals typically have specific waveform features. By counting the number of rising edges within 30 ns after the first rising edge, noise can be further eliminated. Figure \ref{fig:13} shows the detection efficiency as a function of High Voltage, before and after the continuous oscillation check. The working point of the MRPC prototype was selected at 7680 V. At this working point, by applying the continuous oscillation check, the NTR was reduced from 0.9 Hz to 0.004 Hz, while the loss rate of true signals remained below 10\textsuperscript{-4}. The noise shown in figure \ref{fig:7}(c) can be excluded by limiting the baseline Root Mean Square (RMS) value. The baseline RMS distribution of the 8-gap MRPC detecting 0.511 MeV gamma rays is shown in figure \ref{fig:6}(c), with an average baseline RMS value of 3 mV. The noise shown in figure \ref{fig:7}(d) can be eliminated by determining whether the amplitude of the first threshold-crossing peak is the maximum amplitude of the waveform.

\section{Positioning and imaging of $^{22}$Na radioactive source}

\begin{figure}[htbp]
\centering
\includegraphics[width=0.45\textwidth]{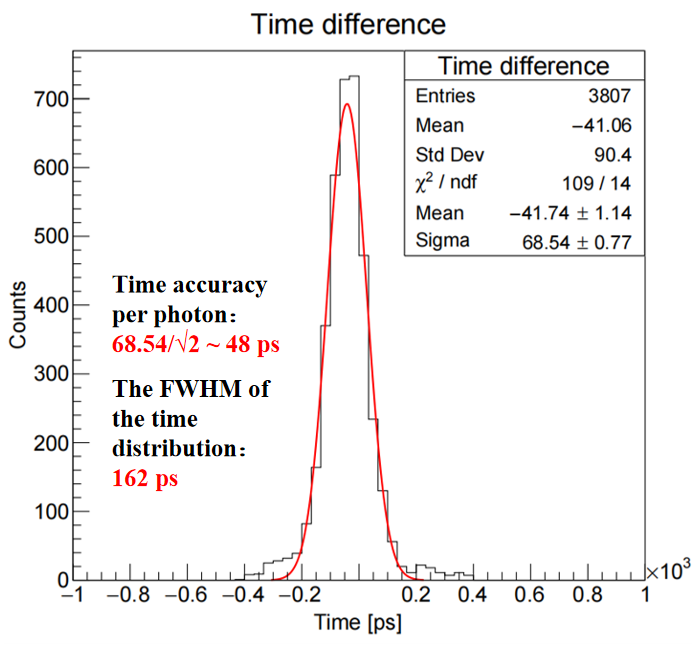}
\caption{Distribution of the time difference in the Gamma test of the 8-gap MRPC detector.\label{fig:9}}
\end{figure}

By measuring the induced signals on adjacent readout strips, and then using the Center of Gravity (COG) method, the position of the gamma interaction point perpendicular to the direction of the readout strip can be determined. The basic principle of the center of gravity method involves calculating the centroid position of the charge distribution by measuring the amplitude and position of each charge signal, thereby determining the position of particles. When an incident gamma photon interacts with the MRPC detector, it generates signals of varying amplitudes on different readout strips of the detector. The amplitudes of these signals correlate with the interaction position of the particle and the MRPC detector, forming a charge distribution. Based on the signal amplitude of each readout strip of the MRPC detector and their corresponding positional coordinates, the center of gravity of the charge is calculated using a weighted average method. The formula is as follows:
\begin{equation}
\label{eq:2}
\begin{aligned}
x_c=\frac{\sum_i Q_i \cdot x_i}{\sum_i Q_i}
\end{aligned}
\end{equation}
where $x_c$ is the position of the center of gravity of the charge, $Q_i$ is the signal amplitude of the $i$-th readout strip, $x_i$ is the spatial position of the $i$-th readout strip, and $\sum_i Q_i$ is the total signal sum from all readout strips. By substituting the coordinates and signal amplitudes corresponding to each readout strip into the equation, the position of the gamma interaction point perpendicular to the direction of the readout strip can be calculated. The position of the gamma interaction point along the length direction of the readout strip is determined according to the time difference of the signals exceeding the threshold at both ends of the readout strip. After determining the positions of the interaction points of the gamma with the MRPC-C and MRPC-B detectors, a response line is drawn. Subsequently, the specific position on this response line is determined according to the time difference of this event between the two detectors. The Full Width at Half Maximum (FWHM) value of the time distribution calculated from the time difference between the two detectors is 162 ps, as shown in figure \ref{fig:9}.

\begin{figure}[htbp]
\centering
\subfigure[]{
\label{fig：subfig_p}
    \includegraphics[width=.4\textwidth]{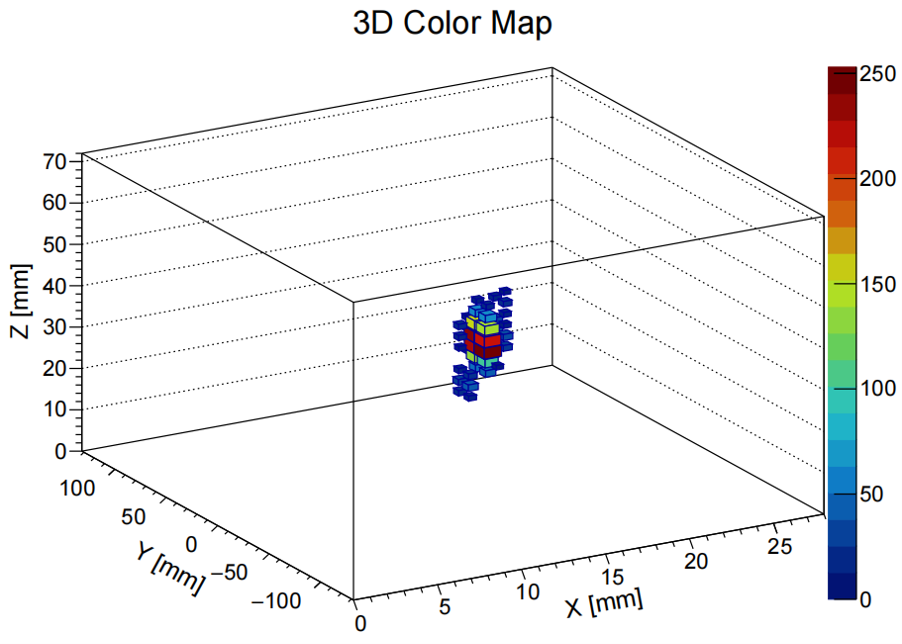}
}
\subfigure[]{
\label{fig：subfig_q}
\includegraphics[width=.38\textwidth]{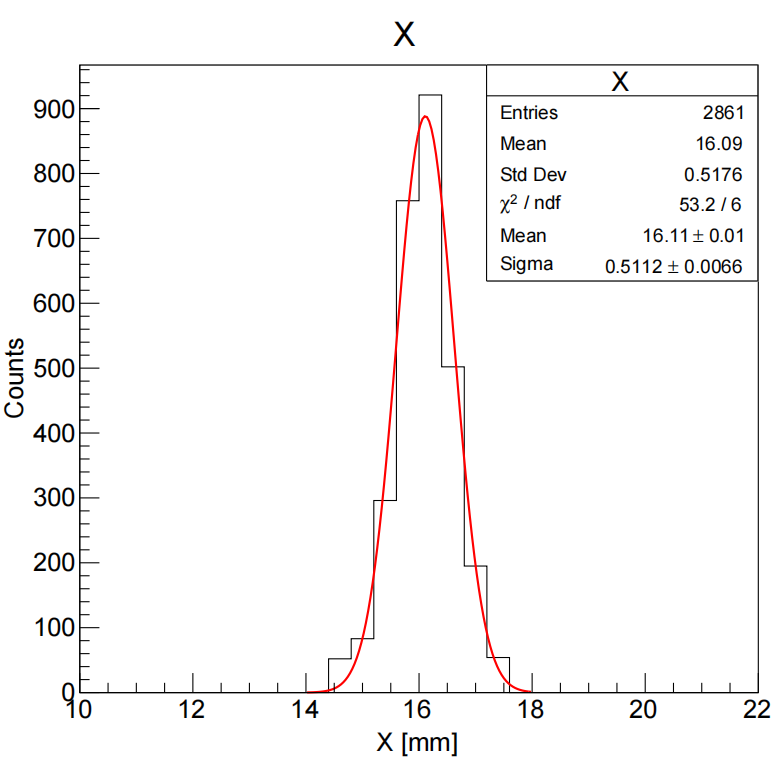}
}
\subfigure[]{
\label{fig：subfig_r}
    \includegraphics[width=.38\textwidth]{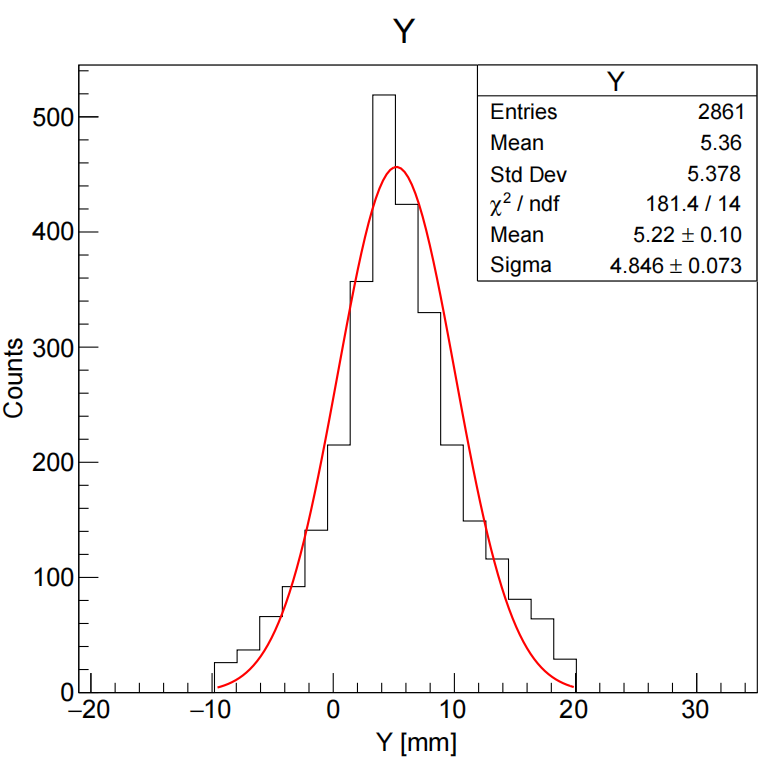}
}
\subfigure[]{
\label{fig：subfig_s}
    \includegraphics[width=.38\textwidth]{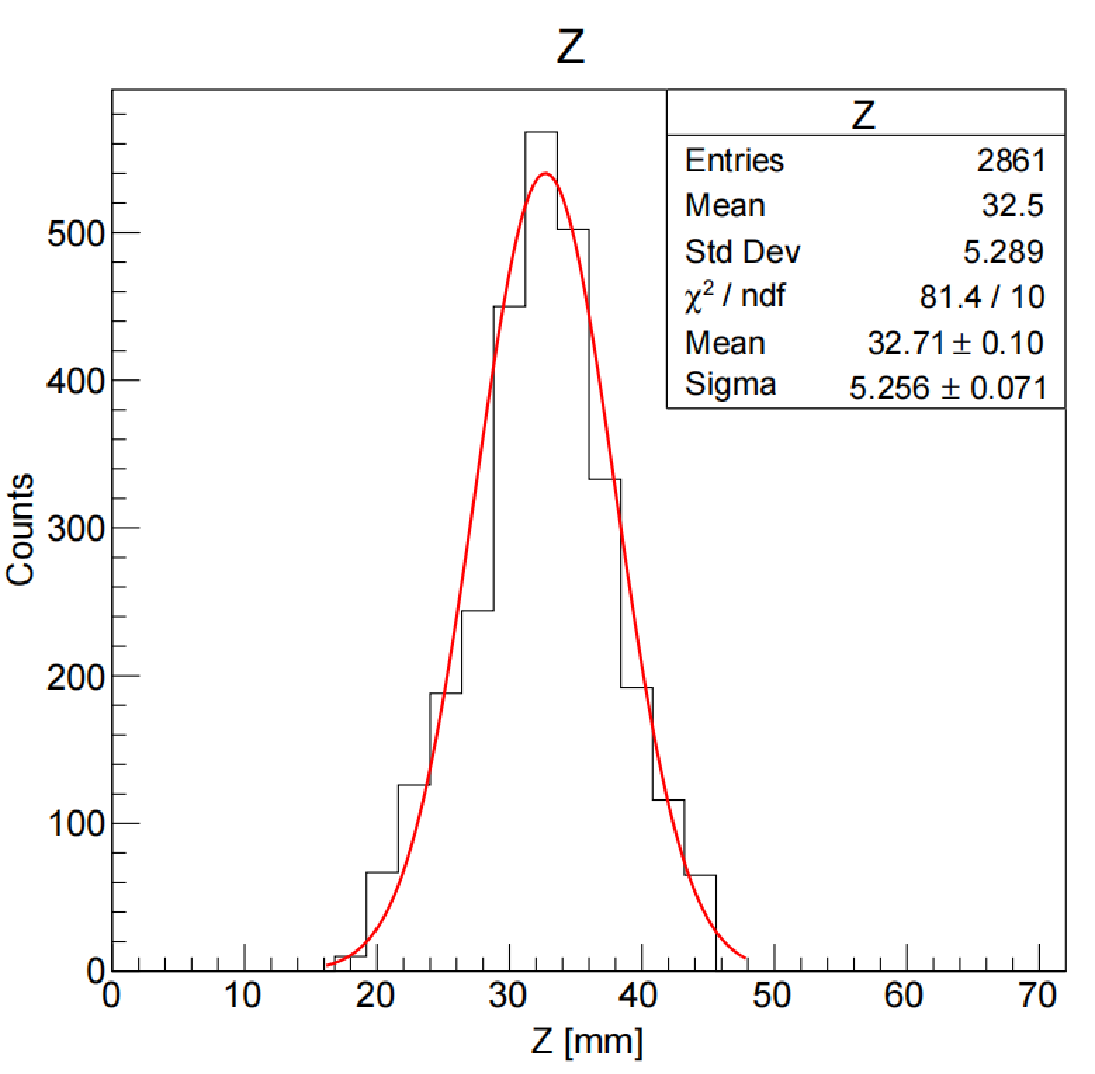}
}
\caption{(a) Three-dimensional image of the $^{22}$Na source positioned by TOF and the projections of each positioning point in (b) the X-axis, (c) the Y-axis, and (d) the Z-axis directions. \label{fig:10}}
\end{figure}

The positioning points are presented as a three-dimensional image, as shown in figure \ref{fig:10}(a). The X, Y, and Z axes are divided into 30 bins each. A color bar is used to represent the magnitude of the bin content values, with the color range dynamically scaled from the minimum to the maximum bin content values. To enhance the visualization of high-density regions, a threshold is applied to the bin content values, set to 4\% of the maximum intensity on the color bar. This results in a 3D image of the localization point distribution, where brown-red regions indicate the areas with the highest concentration of positioning points. The projections of the positioning points in the X-axis, Y-axis, and Z-axis directions are shown in figure \ref{fig:10}(b), (c), and (d) respectively. According to the Time-of-Flight (TOF) method, the coordinates of the $^{22}$Na source are located at (16.11 mm, 5.22 mm, 32.71 mm), which closely match the manually placed source coordinates of (16 mm, 4.5 mm, 32 mm). The $\sigma$ values of the positioning point distributions along the X, Y, and Z axes are 0.51 mm, 4.85 mm, and 5.26 mm, respectively, corresponding to spatial resolutions (FWHM) of 1.20 mm, 11.42 mm, and 12.39 mm. 
\begin{figure}[h]
\centering
\includegraphics[width=0.3\textwidth]{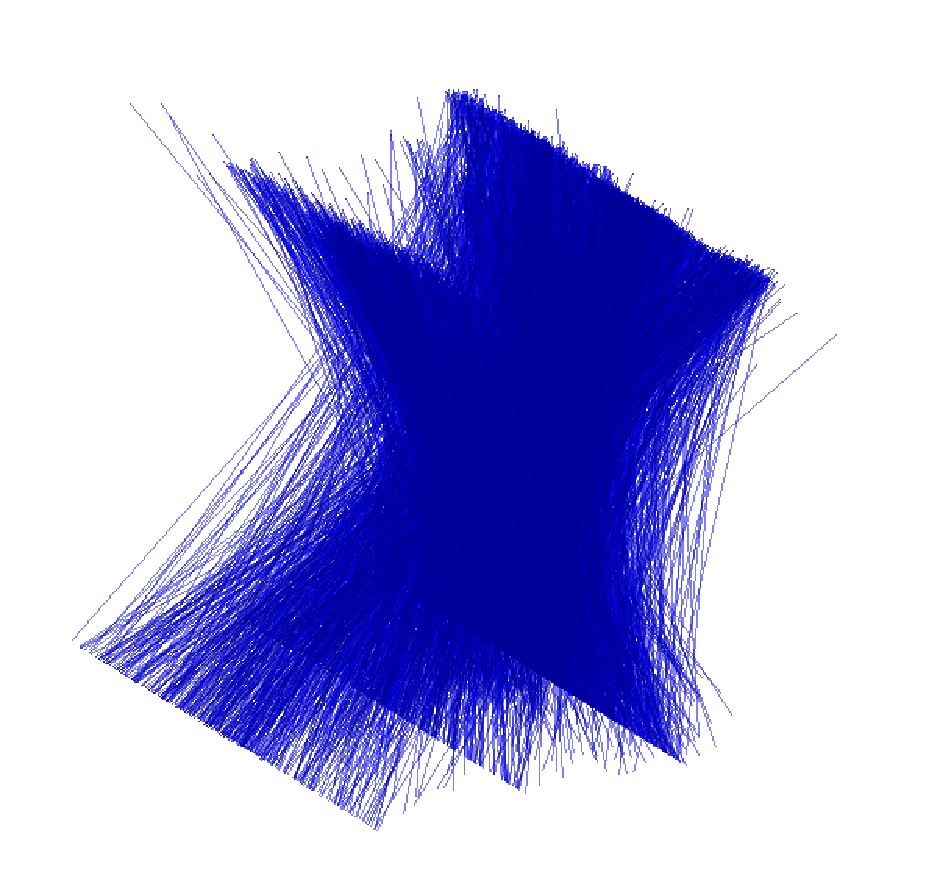}
\caption{Distribution of response lines.\label{fig:11}}
\end{figure}

\begin{figure}[htbp]
\centering
\subfigure[]{
\label{fig：subfig_t}
    \includegraphics[width=.45\textwidth]{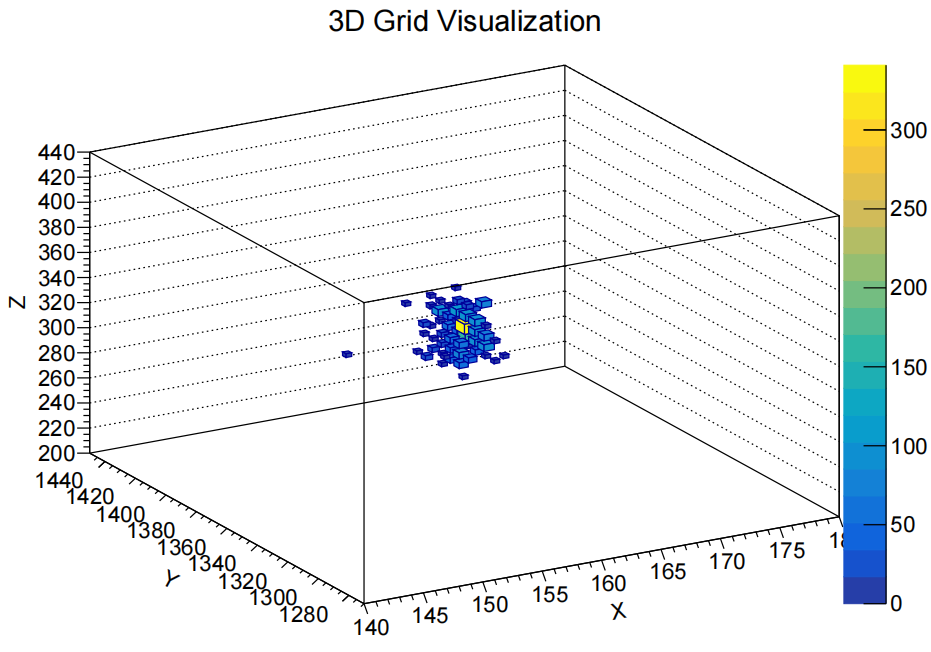}
}
\subfigure[]{
\label{fig：subfig_u}
\includegraphics[width=.4\textwidth]{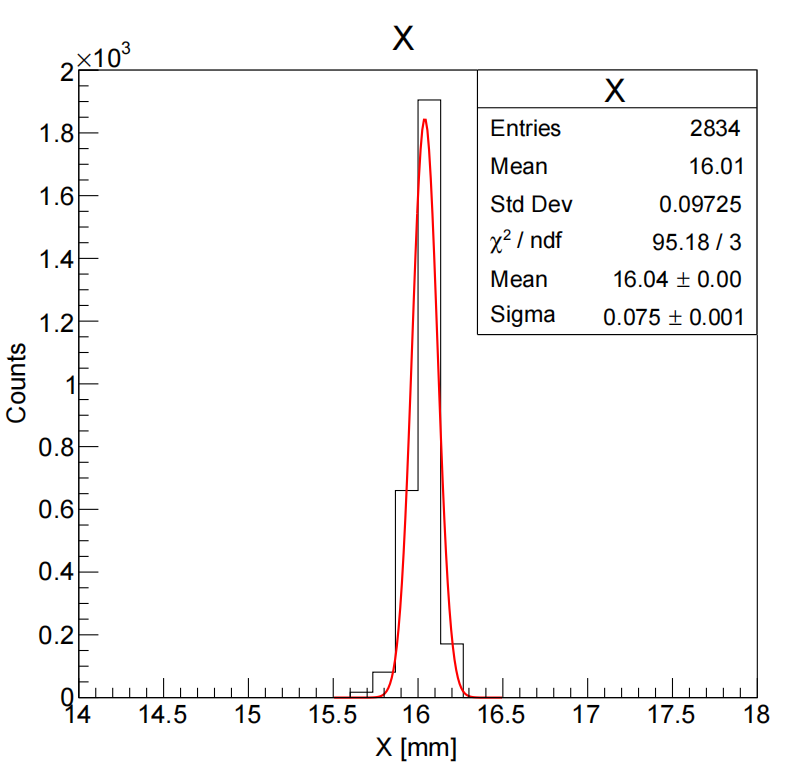}
}
\subfigure[]{
\label{fig：subfig_v}
    \includegraphics[width=.4\textwidth]{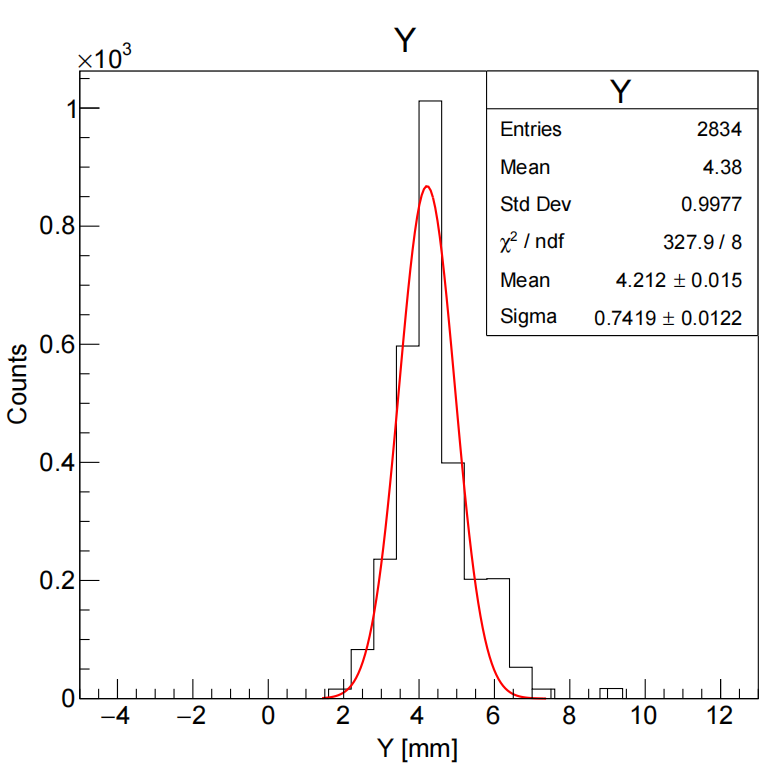}
}
\subfigure[]{
\label{fig：subfig_w}
    \includegraphics[width=.4\textwidth]{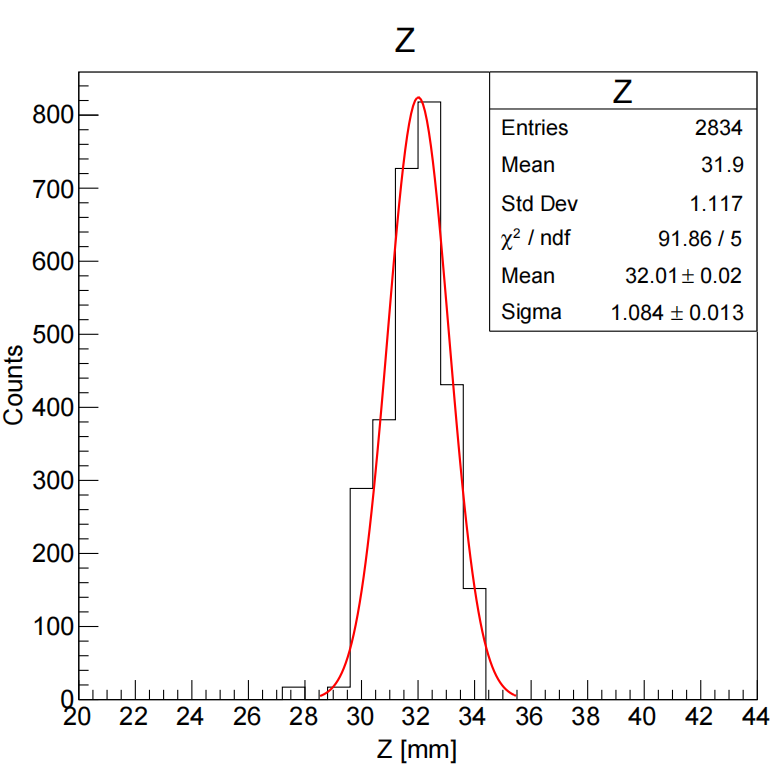}
}
\caption{(a)Three-dimensional pixel map of the intersection points of the response lines and the projection distributions of the intersection points of the response lines in (b) the X-axis, (c) the Y-axis, and (d) the Z-axis directions, which are obtained by conversion from the three-dimensional pixel map. \label{fig:12}}
\end{figure}

In addition, we also positioned the $^{22}$Na source based on the intersection points of the response lines. After determining the positions of the interaction points of gamma photons with the MRPC-C and MRPC-B detectors, a response line is formed by connecting these points. The distribution of the response lines is shown in figure \ref{fig:11}. A three-dimensional pixel map of the intersection points of the response lines is drawn, as shown in figure \ref{fig:12}(a). The yellow color represents the area where the intersection points of the response lines are the most intensive. Ten pixel points are set per 1 mm. The projections of the distribution of the intersection points of the response lines in the X-axis, Y-axis, and Z-axis directions are obtained by conversion according to the three-dimensional pixel map, as shown in figure \ref{fig:12}(b), (c), and (d). The position coordinates of the $^{22}$Na source positioned according to the intersection points of the response lines are (16.04 mm, 4.21 mm, 32.01 mm), and the $\sigma$ values of the distributions in the X-axis, Y-axis, and Z-axis directions are 0.08 mm, 0.74 mm, and 1.08 mm respectively.

\section{Conclusion}
In this study, a self-triggered data acquisition system and an optimized noise reduction algorithm were developed for MRPC TOF-PET systems. The integration of a fast front-end amplifier and a DRS4 waveform digitization module enabled high-speed, high-precision signal processing. A multi-stage noise reduction strategy combining threshold discrimination, coincidence logic, and continuous oscillation check reduced the noise trigger rate to an ultra-low level of 0.004 Hz. The system demonstrated an excellent time resolution of 162 ps FWHM for 0.511 MeV gammas, enabling accurate results for both the localization and imaging of the $^{22}$Na radioactive source. 

\acknowledgments

The work is supported by the National Natural Science Foundation of China under Grant No. 11927901, 11420101004, 11461141011, 11275108, 11735009 and U1832118. This work is also supported by the Ministry of Science and Technology under Grant No. 2020YFE0202001, 2018YFE0205200 and 2016YFA0400100.

% Bibliography

 \bibliographystyle{JHEP}
 \bibliography{reference}

\end{document}